# Nature of low-temperature dense ice up to 80 GPa observed by x-ray diffraction

Hiroshi Fukui[1*], Hirokazu Kadobayashi[2], Hirotaka Abe[3], Ryunosuke Takahashi[3], Hiroki Wadati[1,4], Naohisa Hirao[5]

[1] Graduate School of Science, University of Hyogo, Kouto 3-2-1, Kamigori, Hyogo 678-1297, Japan.
[2] National Institute for Materials Science, Tsukuba, Ibaraki 305-0044, Japan.
[3] Graduate School of Material Science, University of Hyogo, Kouto 3-2-1, Kamigori, Hyogo 678-1297, Japan.
[4] Institute of Laser Engineering, Osaka University, Suita, Osaka 565–0871, Japan
[5] Japan Synchrotron Radiation Research Institute, Kouto 1-1-1, Sayo, Hyogo 679-5198, Japan.
* fukuih@sci.u-hyogo.ac.jp

## Abstract
We have measured the lattice volume of ice VIII in different pressure-temperature pathways and found isothermal compression at low-temperature conditions makes the volume larger. Ice VIII has become its high-pressure phase with the molar volume of 6.45 cm$^3$ at 10 K where the pressure can be estimated as 60.4 GPa based on the third-order Birch-Murnaghan equation with parameters determined in this study ($K_0$= 32.4 GPa, $K_0$'=3.7, and $V_0$= 11.9 cm$^3$). The present results indicate that this high-pressure state is paraelectric with tetragonal symmetry.

## Introduction
Water ice is one of the most familiar materials to us. Meanwhile, this material shows various anomalous properties, mainly due to the polarization of H$_2$O molecules. An appearing phase sometimes depends on pressure/temperature history especially at low-pressure and low-temperature conditions (ices Ih and Ic [e.g., 1], amorphous ices [2,3], etc.).

At high pressures above ca. 2 GPa, ice has high-density packing structures consisting of interpenetrating diamond-like sublattices [4] (Fig. 1), in which each oxygen atom is surrounded by eight nearest neighbors but is connected to only tetrahedrally coordinated four neighbors with hydrogen bonds. Ices with a body-centered-like arrangement of oxygen atoms are called dense ices; they are ice VII, ice VIII, and ice X. Ice VIII has a tetragonal lattice, where the polarization of H$_2$O molecules is along the crystallographic *c* axis. The dense ice contains two sublattices each of which has an ice-Ih-like structure. In ice VIII, each sublattice contains H$_2$O molecules having the same polarization and two sublattices are antiferroelectrically ordered and shift from each other along the crystallographic *c* direction (Fig. 1). That means the distance at the oxygen (δ−) side is shorter than that at the hydrogen (δ+) side of the molecule. The polarization disappears in ice VII, where two hydrogen atoms randomly (or disorderly) occupy two out of four sites around oxygen to satisfy ice rules. Further compression, ice VII and ice VIII transform to ice X, where oxygen atoms form bcc lattice and hydrogen atoms take the center of two oxygen atoms, roughly speaking. The compressibility measurements and theoretical calculation revealed that ice VII' (or sometimes called dynamically disordered ice X) exists between disordered ice VII and ordered ice X [5,6]. The transition between ices VIII and X was confirmed by Raman spectroscopy under isothermal compression/decompression [5]. However, this transition from VIII to X was not yet conformed with respect to their crystallographic structures.

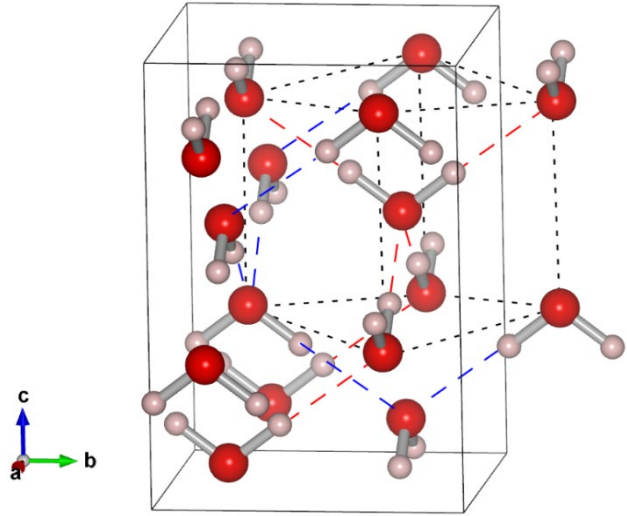

Figure 1. The crystallographic structures of ice VIII based on the structural parameter given in the literature [4] using VESTA [29]. Red and pink spheres indicate oxygen and hydrogen atoms, respectively. Black solid lines indicate the unit cell of ice VIII. H$_2$O molecules with the same polarization are connected with hydrogen bonding and make one sublattice. Black dotted lines indicate the corresponding unit cell for cubic phases (ices VII and X). Red and blue dashed lines indicate hydrogen bonds belonging to different ice-Ic-like sublattices.

Ice VII was reported to show anomalous compression behavior, where the cubic structure deforms to tetragonal under compression at room temperature [7]. As this deformation was not always observed (e.g. Ref. 8), this is probably due to the stress condition where ice VII was being compressed. This fact indicates that ice VII is sensitive to pressure/temperature history. It deduces that ice VIII is so, too. However, as described, the phase transition between ices VIII and X was experimentally confirmed under isothermal conditions and spectroscopic techniques.

We have therefore performed x-ray diffraction as well as Raman spectroscopy measurements on ice VIII passing different pressure/temperature pathways.

**Results and Discussion**

As widely accepted, the tetragonal ice VIII transforms to the cubic ones, ice VII or VII' on heating or ice X on compression. Indications of the transition should be the disappearance of Raman peaks at a low-frequency region, which are translational and rotational modes, the disappearance of some reflections like 101 in XRD, and uniting of 112 and 200 reflections of the tetragonal lattice ($112_T$ and $200_T$) to 110 of the cubic ($110_C$). We have checked the changes of these features. We have performed three runs tracking different pressure and temperature paths.

Figure 2 shows one-dimensional diffraction patterns obtained at 10 and 120 K in Run 3. In ice VIII, oxygen atoms occupy the 8e positions of spacegroup $I4_1/amd$. For the incident x-ray energy used in this study (~30 keV), the form factor of oxygen is approximated to be a real number and that of hydrogen is negligible compared to that of oxygen. In this situation, the intensity of the 101 line is zero when the internal coordinate $z$ of the oxygen atom, $z(O)$, be 1/8. The disappearance of the 101 line does not necessarily mean the transition to the cubic phase. In this condition of $z(O) = 1/8$, the eight distances between oxygen atoms are identical and antiferroelectricity has most probably disappeared. This supports the paraelectricity of the high-pressure phase. Supplementary Figure S1 shows the pressure-temperature conditions where the 101 line was observed or not.

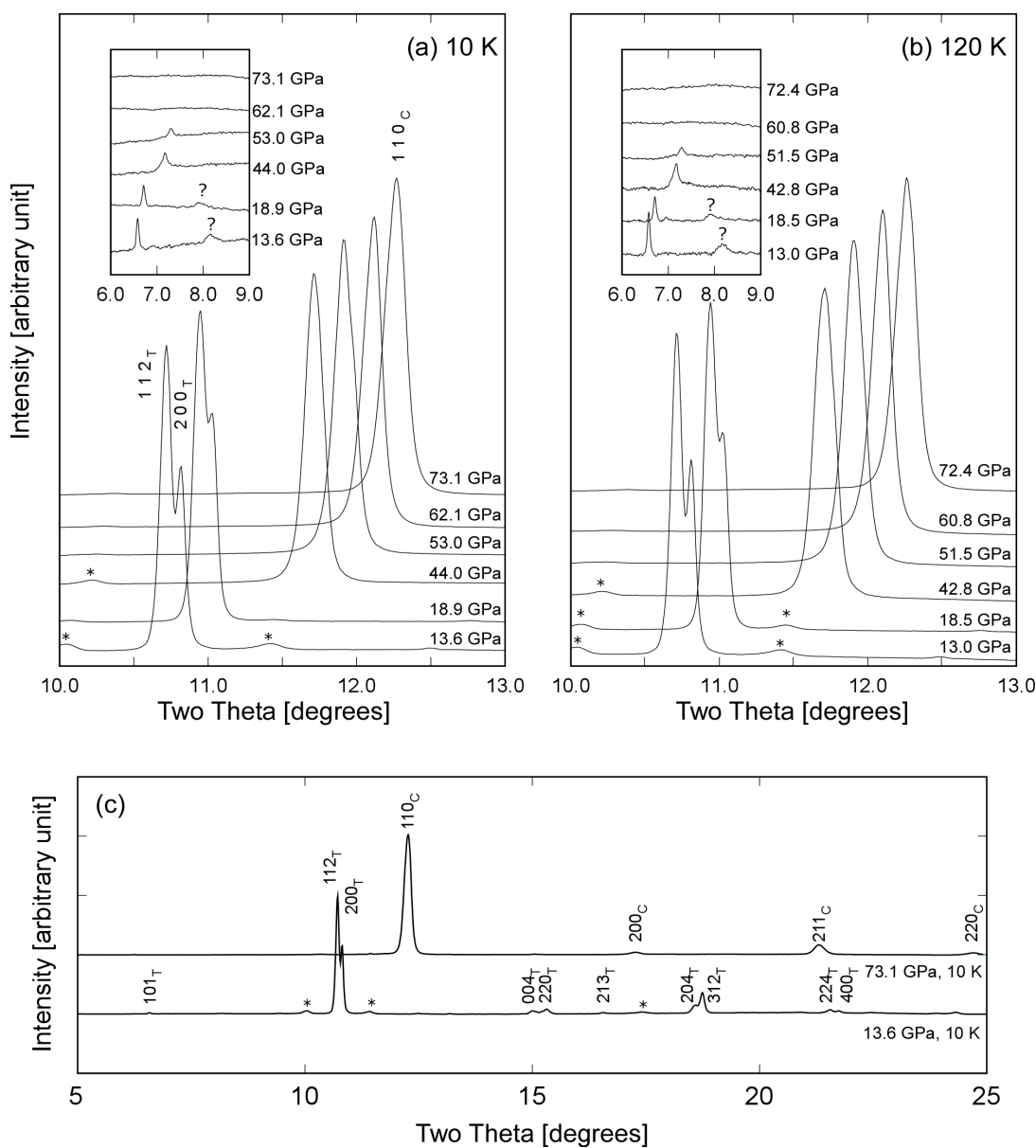

Figure 2. Integrated profiles of the highest peak(s), corresponding the 220 and 112 lines of ice VIII (with subscript T) or the 110 of ices VII or X (with subscript C) (a) at 10 K and (b) 120 K. (inset) those at two theta where the 101 line for ice VIII appears. Backgrounds were subtracted and peak heights are normalized. Peaks indicated by an asterisk (*) are from the rhenium gasket. The peak at around 8 degrees is unidentified. (c) Whole diffraction patterns at 13.6 and 73.1 GPa at 10 K where backgrounds are subtracted. The wavelength was 0.41358 Å.

Note that the $c/\sqrt{2}a$ ratio is not necessarily unity though the nearest O-O distances of intra- and inter-sublattice are the same when $z(O)$ is 1/8. Therefore, one cannot determine the crystallographic symmetry of dense ice based on the appearance/disappearance of the 101 diffraction line ($101_T$) but should do based on the peak separation of the $112_T/200_T$ lines of the tetragonal phase corresponding to the $110_C$ line. However, it is not so straightforward for those obtained at high-pressure conditions. The most intense peak at around two-theta of 11.5 degrees ($112_T$ and $200_T$) looks like a single peak above 40 GPa (Fig. 2). It is difficult to identify a shoulder even when $101_T$ can be observed. Therefore, it is probably natural to consider that the ice VIII transforms to a high-pressure phase when $101_T$ disappears.

To obtain the lattice parameters of dense ice from XRD patterns, one had to consider the effect of uniaxial stress components based on, so-called Gamma plot, as discussed in previous studies for dense ice [5,9]. However, the elastic compliances and the shear modulus under high-pressure conditions are not well determined. Additionally, this relationship cannot be adopted to a material with a tetragonal symmetry lattice. Therefore, the lattice parameters and volumes were calculated by fitting the relation $1/d(hkl) = \sqrt{(h^2+k^2)/a^2 + (l/c)^2}$ to observed 101, 112, 200, 004, 220, 204, and 312 lines. Those for a cubic symmetry lattice were determined based on the 110, 200, 211, and 220 lines. Just for reference, those for the tetragonal symmetry lattice at conditions where $101_T$ was not observed were also determined using the other six lines without $101_T$. The results are summarized in Supplementary Table S1.

The Raman spectra were collected as shown in Supplementary Fig. S2. At low-temperature conditions, two or three peaks were observed. These peak positions shifted to a higher frequency with increasing pressure. At certain pressure conditions, these peaks were disappeared. These results are consistent with previous studies [5]. Four peaks can be attributed to $T_z(A_{1g})+T_{xy}(E_g)$, $T_z(B_{1g})+T_{xy}(E_g)$, $R_{xy}(E_g)$, and $R_z(B_{2g})$ modes from the lower frequency [5]. The peaks corresponding to $T_z(A_{1g})+T_{xy}(E_g)$ and $T_z(B_{1g})+T_{xy}(E_g)$ became doublets at some high-pressure conditions (Supplementary Fig. S2c). This was also reported in the literature. This splitting was observed in Run1. In contrast, the splitting is not so clear at similar conditions in Run 3 (Supplementary Fig. S2c). A noticeable point is that the Raman frequencies in Run 1 are lower than those in Runs 2 and 3 at similar pressure conditions. This is consistent with the discussion on molar volumes as discussed later.

Figure 3 shows the pressure variation of the $c/\sqrt{2}a$ value. They are rather scattered in Run 1 and look monotonically decreasing in Runs 2 and 3. Pruzan *et al.* reported that the $c/a$ ratio exhibited a plateau around 16 GPa at 220 K [5]. Yamawaki *et al.* also observed it above 20 GPa at 87 K (unpublished data) [10]. Both of them are scattered, which is similar to those in Run 1 of the present study. All of the data showing the $c/a$ plateau were measured in low-temperature compression pathways. These results indicate that the $c/a$ ratio monotonically decreases with pressure and this decreasing is somehow suppressed in isothermal compression at low temperature. In contrast, Komatsu *et al.* reported extremely small values, in particular at around 10 GPa [11]. The ice VIII they measured was synthesized in a rapid cooling (6 K/min). They argued that this low $c/a$ ratio was because of the low mobility of hydrogen (deuterium, to be exact) around this pressure. An annealed data gives a similar value of the $c/\sqrt{2}a$ value to those in the present study (Fig. 3) [11].

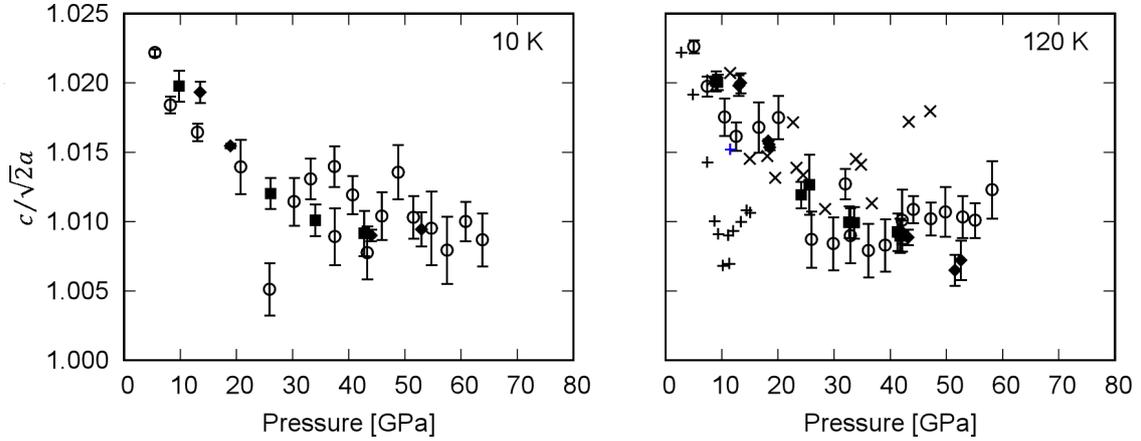

Figure 3. Pressure dependence of the $c/\sqrt{2}a$ ratio. Open circles, closed squares, and closed diamonds indicate the results of Run 1, 2, and 3, respectively. Crosses are based on Ref. 10 (unpublished data). Pluses are from Ref. 11 (the blue one is the value for an annealed sample).

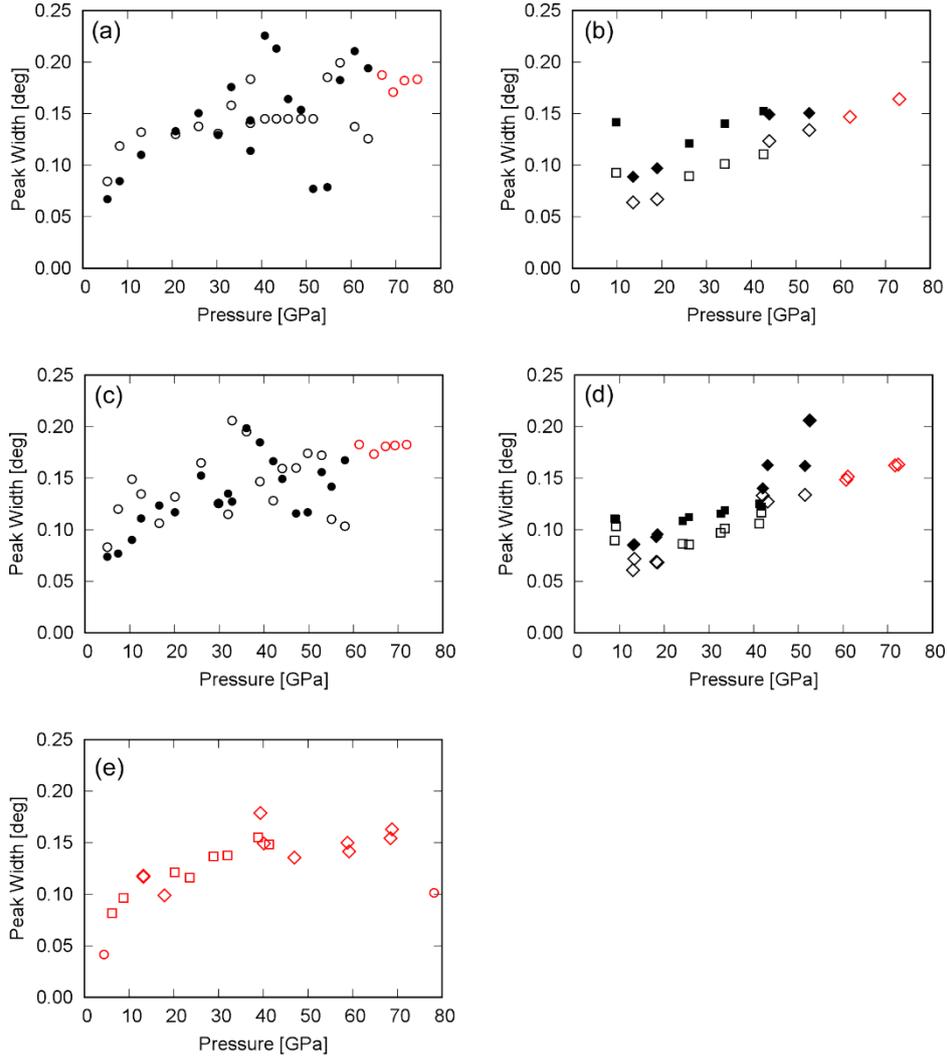

Figure 4. Pressure dependence of peak widths at (a,b) 10 K, (c,d) 120 K, and (e) 300K. Circles, squares, and diamonds are for Run 1, 2, and 3, respectively. Solid, open, and red symbols indicate $112_T$, $200_T$, and $110_C$, respectively.

Figure 4 shows pressure variation of peak width for $112_T$, $200_T$, and $110_C$. The widths for $112_T$ and $200_T$ are rather scattered probably because the present analysis treated not peak intensities but peak positions. However, this comparison helps us understand the differences between Run1 and Runs 2 and 3. Generally, the widths become broad with increasing pressure. The widths are narrower in Runs 2 and 3 than in Run 1 at low temperatures. Although the peak width is not always related only to deviatoric stress, the present data imply that the hydrostaticity was worse in Run 1 than in Runs 2 and 3. It is notable that the width of $110_C$ at 78.2 GPa and 300 K in Run 1 is narrower than those at 40-70 GPa in Runs 2 and 3. This is because the data at 78.2 GPa and 300 K were obtained in a couple of hours after reaching 300 K, whereas data at 300

K in Runs 2 and 3 were collected in a couple of ten minutes after reaching 300 K. This implies that few-hour annealing at 300 K is effective to reduce deviatoric stress.

Obtained molar volumes of ices are plotted in Fig. 5. Note that the difference between molar volumes assuming the cubic and tetragonal symmetries for the high-pressure phase is negligible (Supplementary Table S1). The volumes at around 10 K and around 120 K are larger in pathway 1 than those in pathway 2, whereas those at room temperature are consistent. These facts indicate that the lattice of dense ice was somewhat annealed at room temperature and a stressed lattice has a larger volume than an annealed one at the same pressure and temperature condition. The present results are compared with those in the literature. Yamawaki *et al.* measured the lattice volume of ice VIII compressed at 87 K [10]. Recent neutron scattering studies reported the lattice volumes at 93 and 196 K [11,12]. A theoretical study reported ground state volume of ice VIII [13]. Let us compare them with the present results in Runs 2 and 3. The volumes at 87 K are consistent with the present ones at 120 K. In contrast, those by the neutron studies at 93 and 196 K are slightly smaller than the present ones at 120 K. This may be because of the difference of the pressure maker. The theoretical results are slightly smaller than those at 10 K but reasonably consistent at pressures above 40 GPa. The volumes obtained in Run 1 are larger than all of those except for those below 10 GPa. This probably means that the effect of the cold compression to 10 GPa at 120 K is negligible.

We fit the third-order Birch-Murnaghan equation of state (EoS) [14] to observed volumes of ice VII or VIII. The relationship is as follows:

$P = \frac{3}{2}K_0(x^{-7} - x^{-5})\left\{1 + \frac{3}{4}(K'_0 - 4)(x^{-2} - 1)\right\}$,

where $x = \{V(P)/V_0\}^{1/3}$ and $P$ and $V$ is pressure and volume, respectively. For 300 K, data obtained below 40 GPa were used as a transition from ices VII to VII' occurs at around 40 GPa [6]. For low temperature, data obtained below 60 GPa in Runs 2 and 3 were used because a transition from ice VIII to its high-pressure phase occurs at around 60 GPa as discussed later. Three free parameters, $K_0$, $K'_0$, and $V_0$, have been determined as 30.0(39) GPa, 3.9(2), and 11.8(2) cm$^3$ for 300 K, 21.0(104), 4.4(6), and 12.6(10) cm$^3$ for 120 K, and 32.4(102) GPa, 3.7(4), and 11.9(6) cm$^3$ for 10 K, respectively. The values in parentheses are uncertainties at the last digit. The obtained $V_0$ at 10 K is reasonably consistent with the volume of metastable ice VII at 10 K [12]. Relatively large uncertainties for parameters at low-temperature conditions are probably because of the small number of data. The optimized $V_0$ at 300 K is much smaller than those in previous studies [7, 9, 12, 15, 16]. This is probably because of the lack of low-pressure data. We also fit the equation to the volumes with fixed $V_0$. We used values given in Fig. 6 of Ref. 12, which are mainly based on measurements on recovered dense ice to the ambient pressure. The determined parameters are listed in Table 1.

Table 1. Optimized parameters for the third-order Birch-Murnaghan equation of state.

| Temperature [K] | $K_0$ [GPa] | $K'_0$ | $V_0$ [cm$^3$/mol] | Pressure range |
|:---:|:---:|:---:|:---:|:---|
| 10 | 32.4(102) | 3.7(4) | 11.9(6) | 8-53 GPa |
|  | 30.8(13) | 3.7(1) | 12.029 | 8-53 GPa with fixed $V_0$ |
| 120 | 21.0(104) | 4.4(6) | 12.6(10) | 8-53 GPa |
|  | 26.7(12) | 4.1(1) | 12.129 | 8-53 GPa with fixed $V_0$ |
| 300 | 30.0(39) | 3.9(2) | 11.8(2) | 5-39 GPa |
|  | 17.7(8) | 4.7(2) | 12.717 | 5-39 GPa with fixed $V_0$ |
|  | 4.21 | 7.77 | 14.52 | 5-48 GPa *[7] |
|  | 14.9 | 5.4 | 12.37 | 5-103 GPa [9] |
|  | 13.8 | 5.9 | 12.717 | 2.1-13.7 GPa [12] |
|  | 4.26 | 7.75 | 14.52 | 2.2-170 GPa *[15] |
|  | 23.7 | 4.15 | 12.3 | 4.3-128 GPa [16] |

* Vinet equation [30] was used to obtain the parameters.

The compression curves with these parameters are drawn in Fig. 5. Both parameter sets (free $V_0$ and fixed $V_0$) reasonably reproduce measured data between 10 and 60 GPa. The curves drawn with the fixed-$V_0$ and the free-$V_0$ show difference at very low-pressure conditions (< 10 GPa) at 300 K and but very much consistent at other conditions. The curves are slightly different above 60 GPa at 300 K. This is because the data above 40 GPa were not used for the fitting.

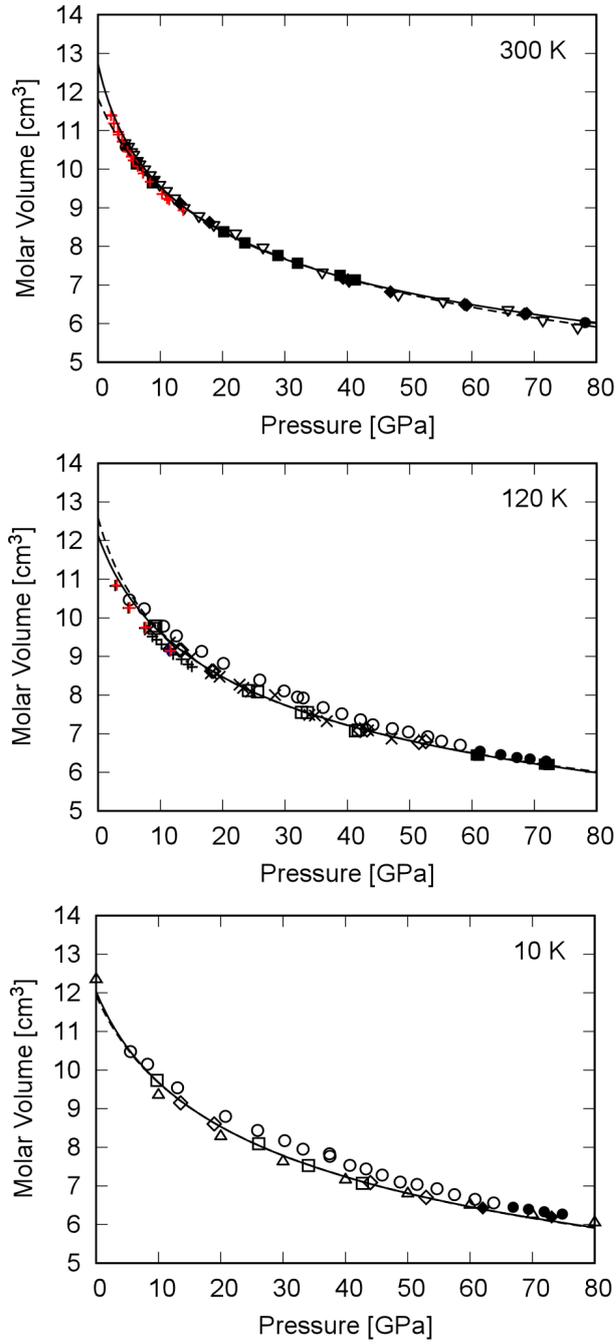

Figure 5. Pressure dependence of molar volume. Circles, squares, and diamonds are for Run 1, 2, and 3, respectively. Open and solid symbols indicate ice VIII and a high-pressure phase, respectively. Crosses are from Ref. 10. Pluses are neutron results (black: Ref. 11, red: Ref. 12). Triangles are values from a theoretical study [13]. Reverse triangles are those from an XRD study [16]. Lines indicate compression curves based on the third-order Birch-Murnaghan equation of state with parameters given in Table 1. Solid and dashed lines are with the $V_0$-fixed parameters and the $V_0$-free parameters, respectively.

The three compression curves and data obtained in Runs 2 and 3 are shown in Supplementary Fig. S3. The temperature dependence on volume at a certain pressure is almost negligible or slightly makes the volume smaller. Calculated volumes at 10, 35, and 60 GPa based on the EoSs are listed in Table 2. Uncertainties on volume were evaluated based on the relationship of $\delta V = \sqrt{\left(\frac{\partial P}{\partial K_0}\right)^2 \delta K_0{}^2 + \left(\frac{\partial P}{\partial K'_0}\right)^2 \delta K'_0{}^2} / \left(\frac{\partial P}{\partial V}\right)^2$, where $\delta X$ means an uncertainty on $X$. Obtained volumes at a certain pressure are almost constant within the uncertainties. If taking them naively, the volume is getting smaller with increasing temperature at pressures where ices VII/VIII is stable and getting larger at the pressure where ice transforms to the high-pressure phase. However, it is not plausible yet. As seen in this study, the cold compression makes the volume of ice VIII larger and annealing at room temperature reduces this effect. It concerns that the annealing at room temperature may not be sufficient to eliminate this effect. The second concern is pressure determination. We used a ruby fluorescence method and a diamond Raman gauge to estimate generated pressure (see Method section). Actually, they worked reasonably well. However, there may be possible systematic errors with temperature. This should be confirmed in some way. Lastly, there seems no physical mechanism for negative thermal expansion in ice VIII. Negative thermal expansivity has been observed in ice I only at low-temperature conditions [17,18]. The mechanism was explained due to negative mode Grüneisen parameters and rotational modes of $H_2O$ molecules [19,20]. Theoretical calculations [21,22] showed translational mode frequencies up to 500 cm$^{-1}$ for ice VIII do not decrease with increasing pressure, implying ice VIII does not show negative thermal expansion. However, the pressure dependence of the librational modes of ice VIII, which corresponds to molecular rotations, should also be confirmed.

Table 2. Molar volumes [cm³/mol] based on the $V_0$-fixed EoS at selected pressure and temperature conditions.

|       | 10 GPa   | 35 GPa    | 60 GPa    |
|-------|----------|-----------|-----------|
| 10 K  | 9.68(7)  | 7.49(10)  | 6.46(12)  |
| 120 K | 9.65(7)  | 7.45(10)  | 6.51(10)  |
| 300 K | 9.55(9)  | 7.40(12)  | 6.49(13)  |

The phase transition to the high-pressure phase occurred between 53.0 and 62.1 GPa at 10 K in Run3 based on the disappearance of 101$_T$. This pressure of 62.1 GPa is consistent with a reported transition pressure of 62 GPa but is meaningfully lower than 63.8 GPa where ice VIII was observed in Run1. The transition in Run 1 occurred between 63.8 and 66.9 GPa at 10 K. This indicates that the cold compression makes the transition pressure higher. Based on the EoS for ice VIII, we estimated pressures from the molar volumes in Run 1. The results indicate that the phase transition occurred between 57.0 (6.561 cm³/mol) and 60.4 GPa (6.450 cm³/mol) at 10 K. This value is still agreed well with the literature value of 62 GPa based on the Raman spectroscopy [5] as shown in Supplementary Fig. S2. However, the transition pressure could be lower because the Raman spectra were measured under cold compression. In the same manner, the phase transition at 120 K occurred between 58.1 (6.703 cm³/mol) and 61.3 GPa (6.546 cm³/mol) in Run 1, corresponding to 53.3 and 58.2 GPa based on the present EoS.

The present study demonstrates the phase transition of ice at a very low-temperature condition based on x-ray diffraction results. The high-pressure phase is most probably paraelectric based on the diffraction patterns and the Raman spectra. The transition pressure determined in cold compression is probably higher than the true pressure (ca. 5 GPa higher in the present case). The transition pressure based on the disappearance of $101_T$ is below 60.4 GPa at 10 K.

## Methods

Distilled, deionized, sterile water from Nippon Gene was used as a sample. A pressure generation device is a diamond anvil cell, which had a pair of diamonds with a 0.3-mm culet. The sample was loaded to a drilled hole on a rhenium gasket with a few ruby tips as pressure markers and compressed to ca. 1 GPa. The DAC was mounted on a closed-cycle helium refrigerator [24] and cooled down to 250 K. This refrigerator was equipped with a helium-gas-driven membrane to compress the DAC. The sample was compressed to about 3 GPa at this temperature to suppress the crystal growth of ice VIII and to try producing a fine powder sample, and then heated to 300K. The dense ices prepared in this way were used in the following compression experiments.

All of the measurements were performed at BL10XU of SPring-8 [24], where one can obtain x-ray diffraction patterns as well as Raman spectra at the same condition. The incident x-ray energies were tuned to ca. 30 keV. X-ray diffraction patterns were collected either the RIGAKU R-AXIS IV++ image plate detector or the PerkinElmer XRD 0822 digital x-ray flat panel detector. During x-ray exposure, the DAC was being oscillated around one axis by ±5 or 8 degrees to improve the crystallite averaging.

Ruby fluorescence and Raman spectra of the sample and diamond were measured with a 532-nm DPSS LASER probe. We also put a couple of ruby tips at the table side of a diamond (the opposite side of a culet) to check the wavelength of the ruby fluorescence at and 10, 120, and 300 K and the ambient pressure ($\lambda_0$). Generated pressure was estimated from the ruby fluorescence shift [25] from $\lambda_0$ in the literature [26]. The $\lambda_0$s were reasonably consistent with the measured numbers. Above 40 GPa, the intensity of the ruby was sometimes too broad or too weak to determine the pressure precisely. Therefore we used the diamond Raman gauge above 30 GPa [27]. As the diamond gauge was developed to determine the pressure at a megabar (>100GPa) range, the scale [27] is inaccurate below 20 GPa. However, pressures from the ruby [24] and the diamond edge [27] above ~20 GPa are consistent, that has been confirmed in the present study (Supplementary Tables S2 and S3 and Fig. S4). The temperature effect on the diamond Raman gauge is most probably negligible at low-temperature conditions because of the large Debye temperature of diamond (> 2200 K) [28]. The pressure uncertainties determined by the diamond edge in this study do not impact the EoS parameters.

We adopt two experimental pathways in separated three runs (Supplementary Figs. S1 and S2), each of which we used a fresh sample. The first one (Run 1) is cooling the sample down to 10 K, heating it to 120 K, and compressing it to the next pressure. A bunch of measurements was done at around 120 K and 10 K. At the end of this pathway, the sample was heated to room temperature. The second pathway (Runs 2 and 3) is compressing the sample to a target pressure at room temperature, cooling it down to ca. 10 K, and then heating up to room temperature. This was repeated. XRD patterns and Raman spectra were measured at room temperature, 120 K and 10 K. A typical cooling/heating ratio is 2 K/min (never faster than 3 K/min). The temperature stability during measurement at a certain condition was less than 3 K. In Run 1, the pressure given by ruby was getting larger above 40 GPa, implying the ruby was getting to feel deviatoric stress (Supplementary Table S2 and Fig. S4).

**Acknowledgement**

Dr. Hiroshi Yamawaki kindly provided their unpublished data on the *c/a* ratio. The authors thank to Dr. Toshiaki Iitaka for discussion and to Dr. Saori Imada-Kawaguchi, Keisuke Okai, Kodai Miura, and Atsushi Fujita for their helps in the experiments. A series of measurements were performed with the approvals of the Japan Synchrotron Radiation Research Institute (Nos. 2019B1240, 2020A0623, and 2020A1539). This work was partially supported by JSPS KAKENHI to HF (grant numbers JP19H02004) and to HK (19K14815).

**Author contributions**

H.F. designed the present study. All authors contributed the experiments. H.F. and H.K. analyzed the data. H.F. wrote the manuscript. All authors reviewed the manuscript.

Supplementary Information for

**Nature of low-temperature dense ice up to 80 GPa observed by x-ray diffraction**

Hiroshi Fukui[1*], Hirokazu Kadobayashi[2], Hirotaka Abe[3], Ryunosuke Takahashi[3], Hiroki Wadati[1,4], Naohisa Hirao[5]

[1] Graduate School of Science, University of Hyogo, Kouto 3-2-1, Kamigori, Hyogo 678-1297, Japan.
[2] National Institute for Materials Science, Tsukuba, Ibaraki 305-0044, Japan.
[3] Graduate School of Material Science, University of Hyogo, Kouto 3-2-1, Kamigori, Hyogo 678-1297, Japan.
[4] Institute of Laser Engineering, Osaka University, Suita, Osaka 565–0871, Japan
[5] Japan Synchrotron Radiation Research Institute, Kouto 1-1-1, Sayo, Hyogo 679-5198, Japan.

This file contains
Supplementary Table S1
Supplementary Table S2
Supplementary Table S3
Supplementary Figure S1
Supplementary Figure S2
Supplementary Figure S3
Supplementary Figure S4

Supplementary Table S1. Obtained lattice parameters and molar volumes. Values at column cubic were calculated using four diffraction lines (110, 200, 211, and 220) of a cubic cell. Those at column tetragonal were calculated using seven lines (101, 112, 200, 004, 220, 204, and 312) of a tetragonal cell. At conditions where the 101 line of the tetragonal cell was not observed, values were calculated using the other six lines for reference. Values in the parentheses are fitting uncertainties at the last digit.

| | $P$ [GPa] | $T$ [K] | cubic | | tetragonal | | |
|---|---|---|---|---|---|---|---|
| | | | $a$ [Å] | $V$ [cm$^3$/mol] | $a$ [Å] | $c$ [Å] | $V$ [cm$^3$/mol] |
| Run 1 | 4.6 | 300 | 32753(2) | 10.576(3) | - | - | - |
| | 78.2 | 300 | 2.715(2) | 6.02(3) | - | - | - |
| | 5.0 | 120 | - | - | 4.582(1) | 6.626(2) | 10.47(1) |
| | 7.4 | 120 | - | - | 4.552(2) | 6.565(3) | 10.24(2) |
| | 10.6 | 120 | - | - | 4.487(4) | 6.457(7) | 9.78(3) |
| | 12.6 | 120 | - | - | 4.451(3) | 6.396(5) | 9.53(2) |
| | 16.8 | 120 | - | - | 4.387(5) | 6.308(9) | 9.13(4) |
| | 20.4 | 120 | - | - | 4.335(4) | 6.237(8) | 8.82(3) |
| | 26.5 | 120 | - | - | 4.276(5) | 6.100(10) | 8.39(4) |
| | 30.7 | 120 | - | - | 4.227(5) | 6.028(9) | 8.11(4) |
| | 32.0 | 120 | - | - | 4.194(2) | 6.006(5) | 7.95(2) |
| | 32.9 | 120 | - | - | 4.195(5) | 5.986(9) | 7.93(4) |
| | 36.0 | 120 | - | - | 4.152(5) | 5.918(9) | 7.68(4) |
| | 39.1 | 120 | - | - | 4.122(5) | 5.878(9) | 7.52(4) |
| | 42.1 | 120 | - | - | 4.091(6) | 5.844(10) | 7.36(4) |
| | 44.1 | 120 | - | - | 4.066(2) | 5.813(4) | 7.23(2) |
| | 47.3 | 120 | - | - | 4.048(3) | 5.783(5) | 7.13(2) |
| | 49.8 | 120 | - | - | 4.031(4) | 5.761(8) | 7.04(3) |
| | 52.9 | 120 | - | - | 4.009(4) | 5.728(7) | 6.93(3) |
| | 55.1 | 120 | - | - | 3.986(3) | 5.694(6) | 6.81(2) |
| | 58.1 | 120 | - | - | 3.963(5) | 5.673(9) | 6.70(4) |
| | 61.3 | 120 | 2.791(1) | 6.55(2) | 3.936(3) | 5.604(5) | 6.53(2) |
| | 64.6 | 120 | 2.7785(7) | 6.46(1) | 3.922(3) | 5.566(4) | 6.44(2) |
| | 67.2 | 120 | 2.768(1) | 6.38(2) | 3.903(4) | 5.550(8) | 6.36(3) |
| | 69.3 | 120 | 2.7629(7) | 6.35(1) | 3.900(3) | 5.535(5) | 6.33(2) |
| | 71.9 | 120 | 2.7547(5) | 6.292(8) | 3.889(3) | 5.518(5) | 6.28(2) |

| P [GPa] | T [K] | | | a [Å] | | V [cm³/mol] |
|---:|---:|---:|---:|---:|---:|---:|
| 5.5 | 10 | - | - | 4.5841(7) | 6.627(1) | 10.479(6) |
| 8.2 | 10 | - | - | 4.542(2) | 6.542(3) | 10.15(1) |
| 13.1 | 10 | - | - | 4.452(2) | 6.399(3) | 9.54(1) |
| 21.1 | 10 | - | - | 4.337(5) | 6.219(9) | 8.80(4) |
| 26.4 | 10 | - | - | 4.289(5) | 6.096(8) | 8.44(4) |
| 31.1 | 10 | - | - | 4.234(4) | 6.0567 | 8.17(4) |
| 33.8 | 10 | - | - | 4.194(4) | 6.009(7) | 7.96(3) |
| 37.5 | 10 | - | - | 4.171(4) | 5.981(7) | 7.83(3) |
| 37.4 | 10 | - | - | 4.167(5) | 5.945(9) | 7.77(4) |
| 40.7 | 10 | - | - | 4.120(4) | 5.896(6) | 7.53(3) |
| 43.2 | 10 | - | - | 4.109(5) | 5.856(9) | 7.44(4) |
| 45.9 | 10 | - | - | 4.076(4) | 5.825(8) | 7.28(3) |
| 48.8 | 10 | - | - | 4.038(5) | 5.788(9) | 7.10(4) |
| 51.5 | 10 | - | - | 4.031(4) | 5.759(7) | 7.04(3) |
| 54.6 | 10 | - | - | 4.010(6) | 5.725(12) | 6.93(5) |
| 57.5 | 10 | - | - | 3.983(6) | 5.677(11) | 6.78(4) |
| 60.9 | 10 | - | - | 3.956(4) | 5.651(8) | 6.65(3) |
| 63.8 | 10 | - | - | 3.939(5) | 5.619(8) | 6.56(3) |
| 66.9 | 10 | 2.778(2) | 6.45(3) | 3.925(4) | 5.559(7) | 6.44(3) |
| 69.4 | 10 | 2.7700(7) | 6.40(1) | 3.911(3) | 5.548(5) | 6.38(2) |
| 71.9 | 10 | 2.7608(5) | 6.33(1) | 3.897(3) | 5.530(4) | 6.32(2) |
| 74.8 | 10 | 2.7518(7) | 6.27(1) | 3.885(3) | 5.515(5) | 6.26(2) |

Supplementary Table S1 (continued).

| | P [GPa] | T [K] | cubic | | tetragonal | | |
| | | | a [Å] | V [cm³/mol] | a [Å] | c [Å] | V [cm³/mol] |
|---|---:|---:|---:|---:|---:|---:|---:|
| Run 2 | 6.18 | 295 | 3.2295(9) | 10.14(2) | - | - | - |
| | 8.74(2) | 296 | 3.176(2) | 9.64(3) | - | - | - |
| | 20.2(6) | 298 | 3.030(2) | 8.38(4) | - | - | - |
| | 23.6(1) | 297 | 2.995(3) | 8.09(6) | - | - | - |
| | 29.2(3) | 299 | 2.954(4) | 7.76(6) | - | - | - |
| | 31.9(3) | 296 | 2.929(4) | 7.57(6) | - | - | - |
| | 39.1(4) | 297 | 2.887(5) | 7.25(7) | - | - | - |
| | 41.5(2) | 296 | 2.872(4) | 7.13(6) | - | - | - |

| P [GPa] | T [K] | cubic a [Å] | V [cm³/mol] | tetragonal a [Å] | c [Å] | V [cm³/mol] |
|---|---|---|---|---|---|---|
| 9.24 | 120 | - | - | 4.484(2) | 6.468(3) | 9.79(2) |
| 8.97(3) | 120 | - | - | 4.475(2) | 6.456(4) | 9.73(2) |
| 26.0(3) | 118 | - | - | 4.217(6) | 6.040(10) | 8.08(4) |
| 24.5(3) | 122 | - | - | 4.224(2) | 6.045(4) | 8.12(2) |
| 33.7(2) | 121 | - | - | 4.126(3) | 5.893(5) | 7.55(2) |
| 32.8(2) | 121 | - | - | 4.127(3) | 5.894(6) | 7.55(2) |
| 41.3(1) | 118 | - | - | 4.038(3) | 5.763(6) | 7.07(3) |
| 41.5(6) | 121 | - | - | 4.043(3) | 5.769(6) | 7.10(2) |
| 9.78 | 11 | - | - | 4.476(3) | 6.455(6) | 9.73(3) |
| 26.2(3) | 10 | - | - | 4.220(3) | 6.040(5) | 8.09(2) |
| 34.2(1) | 10 | - | - | 4.122(3) | 5.889(5) | 7.53(2) |
| 42.3(15) | 10 | - | - | 4.038(4) | 5.763(7) | 7.07(3) |

Supplementary Table S1 (continued).

| | P [GPa] | T [K] | cubic a [Å] | V [cm³/mol] | tetragonal a [Å] | c [Å] | V [cm³/mol] |
|---|---|---|---|---|---|---|---|
| Run 3 | 13.0(1) | 293 | 3.1166(4) | 9.112(7) | - | - | - |
| | 17.9 | 292 | 3.060(2) | 8.62(3) | - | - | - |
| | 39.4(5) | 300 | 2.878(7) | 7.17(10) | - | - | - |
| | 40.1(7) | 298 | 2.873(5) | 7.14(7) | - | - | - |
| | 46.9(3) | 299 | 2.830(3) | 6.82(5) | - | - | - |
| | 58.6 | 303 | 2.784(3) | 6.50(6) | - | - | - |
| | 58.6 | 301 | 2.782(3) | 6.48(4) | - | - | - |
| | 68.9(1) | 300 | 2.751(4) | 6.27(6) | - | - | - |
| | 68.5(2) | 300 | 2.748(3) | 6.25(5) | - | - | - |
| | 13.0 | 122 | - | - | 4.388(2) | 6.329(5) | 9.17(2) |
| | 13.4 | 119 | - | - | 4.387(2) | 6.328(4) | 9.16(2) |
| | 18.5 | 120 | - | - | 4.3050(7) | 6.182(1) | 8.621(5) |
| | 18.3(1) | 123 | - | - | 4.3042(5) | 6.1833(9) | 8.620(4) |
| | 42.8(4) | 122 | - | - | 4.041(3) | 5.768(5) | 7.09(2) |
| | 42.9(6) | 120 | - | - | 4.044(2) | 5.770(3) | 7.10(1) |
| | 52.6(1) | 120 | - | - | 3.988(4) | 5.680(8) | 6.80(3) |

| | | | | | | |
|---|---|---|---|---|---|---|
| 51.5(2) | 120 | - | - | 3.984(3) | 5.671(6) | 6.77(3) |
| 61.2(3) | 124 | 2.777(4) | 6.45(6) | 3.923(4) | 5.583(8) | 6.47(3) |
| 60.8(1) | 127 | 2.777(4) | 6.45(6) | 3.923(5) | 5.579(8) | 6.46(3) |
| 72.4(1) | 124 | 2.741(4) | 6.20(6) | 3.872(5) | 5.509(10) | 6.21(4) |
| 71.6(1) | 121 | 2.743(4) | 6.21(6) | 3.873(5) | 5.510(9) | 6.22(4) |
| | | | | | | |
| 13.6 | 11 | - | - | 4.386(3) | 6.322(5) | 9.15(2) |
| 18.9 | 11 | - | - | 4.3025(6) | 6.179(1) | 8.607(5) |
| 44.1(2) | 11 | - | - | 4.041(2) | 5.767(2) | 7.09(1) |
| 53.0(1) | 10 | - | - | 3.967(4) | 5.663(7) | 6.71(3) |
| 62.2(2) | 13 | 2.774(4) | 6.43(6) | 3.919(5) | 5.572(9) | 6.44(4) |
| 73.1(1) | 11 | 2.741(4) | 6.20(6) | 3.870(5) | 5.507(10) | 6.21(4) |

Supplementary Table S2. Pressure and temperature conditions where the measurement in Run 1 carried out. Values for ruby of each condition are from different rubies in the sample chamber. Numbers in parentheses were not used to estimate the pressure but just for reference.

|  | $T$[K] | $P$[GPa] | Ruby | | Diamond edge | |
|---|---|---|---|---|---|---|
|  |  |  | λ [nm] | $P$[GPa] | ν [cm$^{-1}$] | $P$[GPa] |
| Run 1 | 300 | 4.6 | 695.94 | 4.55 | - | - |
|  | 300 | 78.2 | - | - | 1494.23 | 78.23 |
|  | 120 | 5.0 | 695.17 | 4.96 | - | - |
|  | 120 | 7.4 | 695.95 | 7.15 | (1358.85 | 11.25) |
|  |  |  | 696.19 | 7.56 |  |  |
|  | 120 | 10.6 | 697.05 | 10.25 | (1367.86 | 15.30) |
|  |  |  | 697.27 | 10.88 |  |  |
|  | 120 | 12.6 | 697.76 | 12.26 | (1369.67 | 16.12) |
|  |  |  | 697.98 | 12.92 |  |  |
|  | 120 | 16.8 | 699.18 | 16.39 | (1377.78 | 19.84) |
|  |  |  | 699.46 | 17.19 |  |  |
|  | 120 | 20.4 | 700.38 | 19.87 | (1381.99 | 21.77) |
|  |  |  | 700.73 | 20.92 |  |  |
|  | 120 | 26.5 | 702.53 | 26.26 | (1391.94 | 26.41) |
|  |  |  | 702.68 | 26.72 |  |  |
|  | 120 | 30.7 | 703.75 | 29.94 | (1397.24 | 29.24) |
|  |  |  | 704.11 | 31.39 |  |  |
|  | 120 | 32.0 | 704.73 | 32.93 | 1401.69 | 31.03 |
|  |  |  | 704.69 | 32.81 |  |  |
|  | 120 | 32.9 | (705.76 | 36.09) | 1405.52 | 32.86 |
|  |  |  | (705.74 | 36.03) |  |  |
|  | 120 | 36.0 | (706.63 | 38.79) | 1412.04 | 36.00 |
|  |  |  | (706.37 | 37.97) |  |  |
|  | 120 | 39.1 | (707.26 | 40.75) | 1418.38 | 39.08 |
|  |  |  | (708.88 | 45.88) |  |  |
|  | 120 | 42.1 | - | - | 1424.46 | 42.07 |
|  | 120 | 44.1 | - | - | 1428.63 | 44.13 |
|  | 120 | 47.3 | - | - | 1434.88 | 47.25 |
|  | 120 | 49.8 | - | - | 1439.98 | 49.81 |

| | | | | | |
|---|---|---|---|---|---|
| 120 | 52.9 | - | - | 1446.04 | 52.88 |
| 120 | 55.1 | - | - | 1450..40 | 55.11 |
| 120 | 58.1 | - | - | 1456.29 | 58.13 |
| 120 | 61.3 | - | - | 1462.44 | 61.32 |
| 120 | 64.6 | - | - | 1468.65 | 64.57 |
| 120 | 67.2 | - | - | 1473.62 | 67.18 |
| 120 | 69.3 | - | - | 1477.70 | 69.34 |
| 120 | 71.9 | - | - | 1482.46 | 71.88 |
| 10 | 5.5 | 695.32 | 5.54 | - | - |
| | | 695.26 | 5.37 | | |
| 10 | 8.2 | 696.22 | 8.07 | (1362.59 | 12.92) |
| | | 696.34 | 8.41 | | |
| 10 | 13.1 | 697.81 | 12.60 | (1372.56 | 17.44) |
| | | 698.15 | 13.58 | | |
| 10 | 21.1 | 700.60 | 20.79 | (1384.29 | 22.91) |
| | | 700.84 | 21.40 | | |
| 10 | 26.4 | 702.47 | 26.26 | (1392.16 | 26.51) |
| | | 702.55 | 26.51 | | |
| 10 | 31.1 | 704.02 | 30.94 | (1398.92 | 29.71) |
| | | 704.14 | 31.31 | | |
| 10 | 33.8 | 705.31 | 34.88 | 1403.87 | 32.06 |
| | | 705.18 | 34.48 | | |
| 10 | 37.5 | 706.69 | 39.17 | 1408.71 | 34.39 |
| | | 706.65 | 39.04 | | |
| 10 | 37.4 | (707.81 | 42.68) | 1414.92 | 37.40 |
| | | (708.24 | 44.02) | | |
| 10 | 40.7 | (708.23 | 44.00) | 1421.74 | 40.73 |
| | | (709.37 | 47.63) | | |
| 10 | 43.2 | (707.86 | 42.64) | 1424.66 | 43.16 |
| 10 | 45.9 | - | - | 1432.12 | 45.87 |
| 10 | 48.8 | - | - | 1437.93 | 48.78 |
| 10 | 51.5 | - | - | 1443.39 | 51.53 |
| 10 | 54.6 | - | - | 1449.44 | 54.62 |
| 10 | 57.5 | - | - | 1455.02 | 57.48 |
| 10 | 60.9 | - | - | 1461.53 | 60.85 |

| 10 | 63.8 | - | - | 1467.26 | 63.84 |
| 10 | 66.9 | - | - | 1473.02 | 66.87 |
| 10 | 69.4 | - | - | 1477.71 | 69.35 |
| 10 | 71.9 | - | - | 1482.53 | 71.92 |
| 10 | 74.8 | - | - | 1487.90 | 74.80 |

Supplementary Table S3 Pressure and temperature conditions where the measurement in Runs 2 and 3 carried out. Values at the upper and lower lines of each condition are before and after the XRD measurement, respectively. Numbers in parentheses were not used to estimate the pressure but just for reference.

|  | $T$ [K] | $P$ [GPa] | Ruby | | Diamond edge | |
|---|---|---|---|---|---|---|
|  |  |  | $\lambda$ [nm] | $P$ [GPa] | $v$ [cm$^{-1}$] | $P$ [GPa] |
| Run 2 | 295 | 6.18 | 696.49 | 6.18 | (1353.96 | 9.08) |
|  | 296 | 8.74 | 697.39 | 8.72 | (1358.14 | 10.94) |
|  |  |  | 697.41 | 8.76 |  |  |
|  | 298 | 20.2 | 701.24 | 19.81 | (1377.26 | 19.59) |
|  |  |  | 701.58 | 20.79 | (1379.05 | 20.41) |
|  | 297 | 23.6 | 702.55 | 23.66 | (1385.60 | 23.45) |
|  |  |  | 702.57 | 23.72 |  |  |
|  | 299 | 29.2 | 704.28 | 28.85 | (1396.16 | 28.40) |
|  |  |  | 704.46 | 29.39 |  |  |
|  | 296 | 31.9 | 705.20 | 31.62 | 1404.12 | 32.18 |
|  |  |  | 705.28 | 31.86 |  |  |
|  | 297 | 39.1 | 707.76 | 39.50 | 1417.58 | 38.69 |
|  |  |  | 707.60 | 39.01 |  |  |
|  | 296 | 41.3 | 708.36 | 41.38 | 1423.24 | 41.47 |
|  |  |  | 708.27 | 41.10 |  |  |
|  | 120 | 9.24 | 696.69 | 9.24 | (1361.96 | 12.46) |
|  | 120 | 8.97 | 696.59 | 8.94 | (1360.80 | 12.12) |
|  |  |  | 696.60 | 8.99 |  |  |
|  | 118 | 26.0 | 702.32 | 25.64 | (1388.66 | 24.87) |
|  |  |  | 702.55 | 26.31 |  |  |
|  | 122 | 24.5 | 702.05 | 24.82 | (1387.02 | 24.11) |
|  |  |  | 701.84 | 24.18 |  |  |
|  | 121 | 33.7 | 705.07 | 33.97 | 1406.81 | 33.48 |
|  |  |  | 704.98 | 33.68 |  |  |
|  | 121 | 32.8 | 704.75 | 32.97 | 1405.06 | 32.64 |
|  |  |  | 704.65 | 32.67 |  |  |
|  | 118 | 41.3 | 707.40 | 41.18 | - | - |
|  |  |  | 707.47 | 41.40 |  |  |
|  | 121 | 41.5 | 707.32 | 40.96 | 1424.79 | 42.23 |

| | | | Ruby | | Diamond edge | |
|---|---|---|---|---|---|---|
| | T [K] | P [GPa] | λ [nm] | P [GPa] | v [cm⁻¹] | P [GPa] |
| | | | 707.41 | 41.24 | | |
| | 11 | 9.78 | 696.85 | 9.86 | (1364.26 | 13.68) |
| | | | 696.79 | 9.69 | | |
| | 10 | 26.2 | 702.34 | 25.90 | (1390.27 | 25.63) |
| | | | 702.57 | 26.56 | | |
| | 10 | 34.2 | 705.08 | 34.19 | 1408.41 | 34.25 |
| | | | 705.13 | 34.35 | 1408.17 | 34.13 |
| | 10 | 42.3 | 707.44 | 41.50 | 1428.83 | 44.23 |
| | | | 707.35 | 41.23 | | |

Supplementary Table S3 (continued). Values at the upper and lower lines of each condition are before and after the XRD measurement, respectively. Numbers with grey color was not used to estimate the pressure but just references.

| | T [K] | P [GPa] | Ruby | | Diamond edge | |
|---|---|---|---|---|---|---|
| | | | λ [nm] | P [GPa] | v [cm⁻¹] | P [GPa] |
| Run 3 | 293 | 13.0 | 698.87 | 13.08 | - | - |
| | | | 699.05 | 12.92 | | |
| | 292 | 17.9 | 700.57 | 17.91 | - | - |
| | 300 | 39.4 | 707.53 | 38.9 | 1420.03 | 39.9 |
| | 298 | 40.1 | 707.92 | 40.01 | 1422.04 | 40.88 |
| | | | 707.77 | 39.54 | | |
| | 299 | 46.9 | 710.00 | 46.57 | 1434.60 | 47.11 |
| | | | 710.12 | 46.95 | | |
| | 303 | 58.6 | (713.90 | 59.1) | 1457.14 | 58.57 |
| | 301 | 58.6 | (714.19 | 60.18) | 1457.12 | 58.56 |
| | | | (714.30 | 60.54) | | |
| | 300 | 68.9 | - | - | 1476.53 | 68.73 |
| | | | | | 1476.99 | 68.97 |
| | 300 | 68.5 | - | - | 1476.38 | 68.65 |
| | | | | | 1475.72 | 68.30 |
| | 122 | 13.0 | 698.03 | 13.04 | (1365.99 | 14.46) |
| | 119 | 13.4 | 698.13 | 13.35 | (1365.93 | 14.43) |
| | 120 | 18.5 | 699.91 | 18.52 | (1374.35 | 18.25) |
| | 123 | 18.3 | 699.84 | 18.30 | (1374.24 | 18.20) |

|  |  |  |  |  |  |
|---|---|---|---|---|---|
|  |  | 699.88 | 18.38 |  |  |
| 122 | 42.8 | 707.77 | 42.36 | 1426.77 | 43.21 |
| 120 | 42.9 | 707.72 | 42.20 | 1427.54 | 43.10 |
|  |  |  |  | 1426.97 | 43.31 |
| 120 | 52.6 | (709.65 | 48.3) | 1445.52 | 52.62 |
|  |  |  |  | 1445.26 | 52.49 |
| 120 | 51.5 | (710.13 | 49.8) | 1442.89 | 51.28 |
|  |  |  |  | 1443.70 | 51.69 |
| 124 | 61.2 | - | - | 1462.64 | 61.42 |
|  |  |  |  | 1461.67 | 60.92 |
| 127 | 60.8 | - | - | 1461.31 | 60.73 |
|  |  |  |  | 1461.54 | 60.85 |
| 124 | 72.4 | - | - | 1483.44 | 72.41 |
|  |  |  |  | 1483.30 | 72.33 |
| 121 | 71.6 | - | - | 1481.99 | 71.63 |
|  |  |  |  | 1482.03 | 71.65 |
| 11 | 13.6 | 698.14 | 13.55 | (1367.82 | 15.28) |
| 11 | 18.9 | 699.99 | 18.94 | (1375.63 | 18.84) |
| 11 | 44.1 | (707.47 | 41.61) | 1428.11 | 43.88 |
|  |  |  |  | 1428.81 | 44.22 |
| 10 | 53.0 | (710.32 | 50.7) | 1445.99 | 52.86 |
|  |  |  |  | 1446.45 | 53.09 |
| 13 | 62.2 | - | - | 1464.44 | 62.36 |
|  |  |  |  | 1463.72 | 61.99 |
| 11 | 73.1 | - | - | 1484.66 | 73.06 |
|  |  |  |  | 1484.79 | 73.13 |

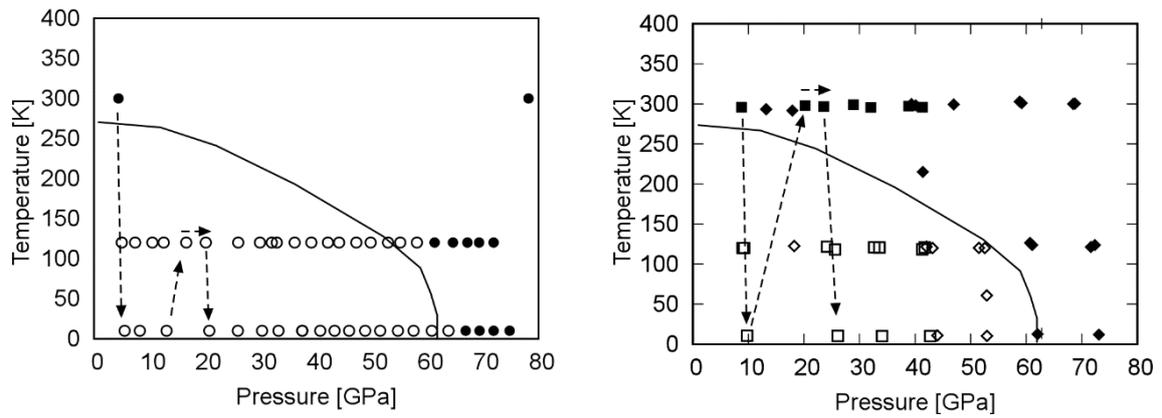

Supplementary Figure S1. Experimental conditions (left: Run1, right: Run2 (squares) and 3 (diamonds)) where the 101 line were observed (open) or not (solid). Some pathways for compression and cooling/heating are also shown by dashed arrows. A reported phase boundary between the ice VII or X and ice VIII by Phruzan et al. [5] is also shown by solid lines.

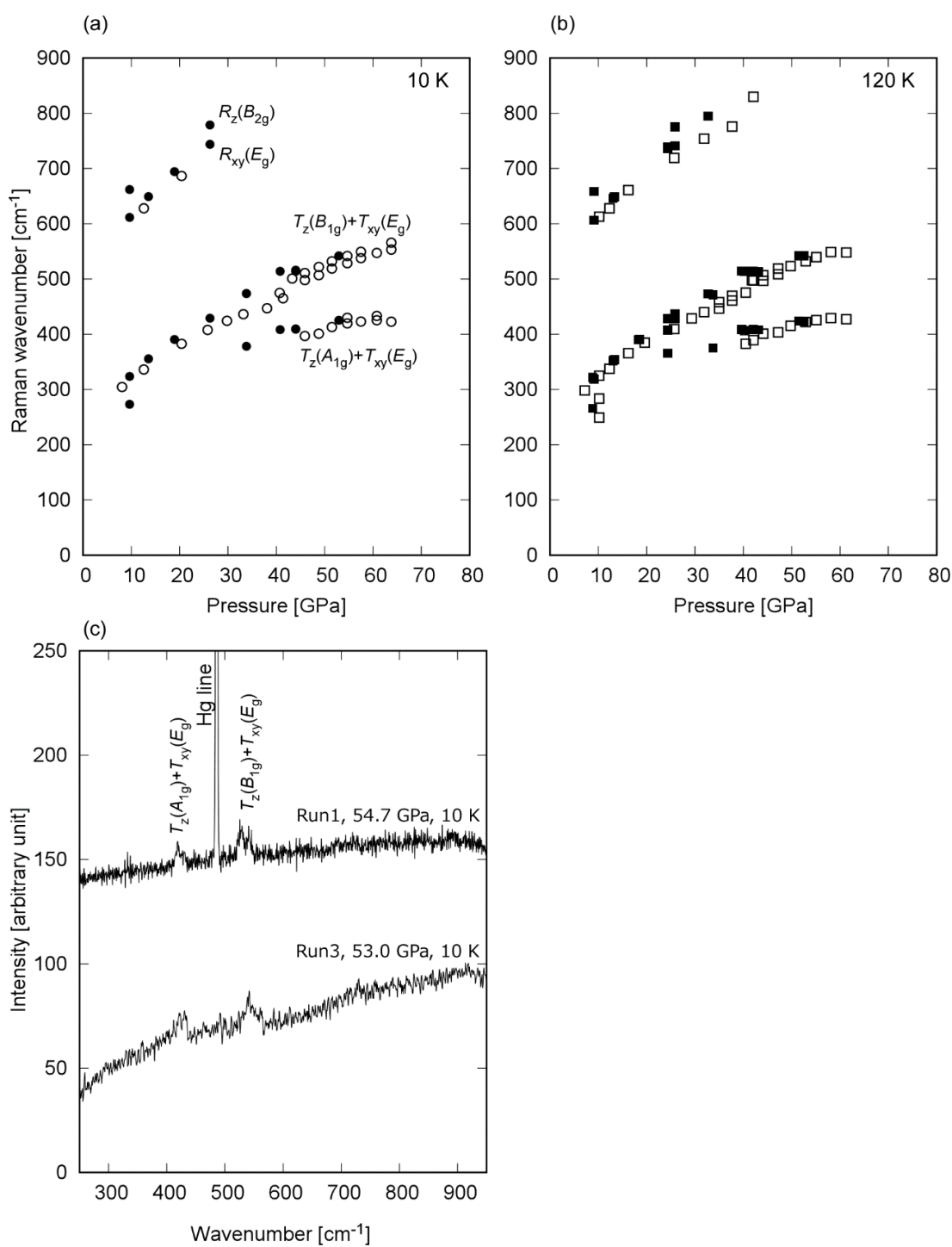

Supplementary Figure S2. Pressure variation of lattice-modes frequencies of ice VIII (a) at 10 K and (b) 120 K. Open and solid symbols are obtained at Run 1 and Runs 2 and 3, respectively. (c) Typical Raman spectra of ice VIII at around 50 GPa and 10 K obtained Runs 1 and 3. A strong line at 484.5 cm$^{-1}$ is mercury e line from the experimental hutch lighting (586.074 nm).

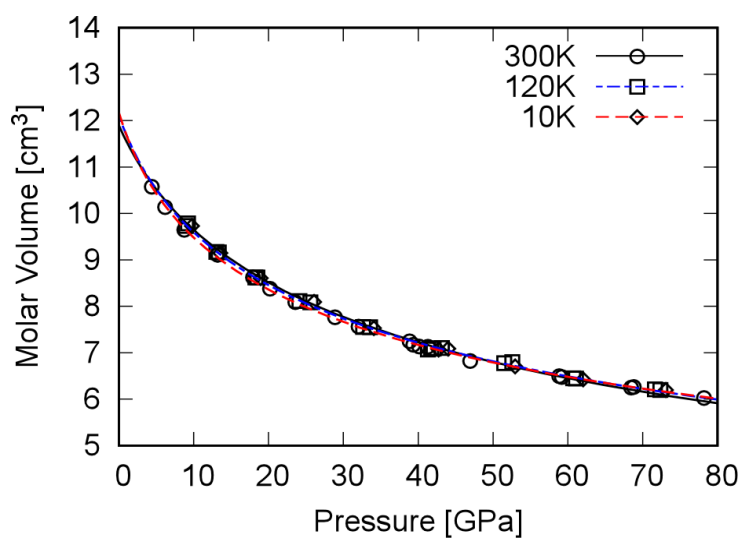

Supplementary Figure S3. Compression curves of dense ice determined with low pressure data (< 40 GPa for 300 K and < 60 GPa for low temperatures). Parameters are shown in Table 1. Solid black, dot-dashed blue, and dashed red lines indicated 300, 120, and 10 K, respectively. Circles, squares, and diamonds are observed data at Runs 2 and 3.

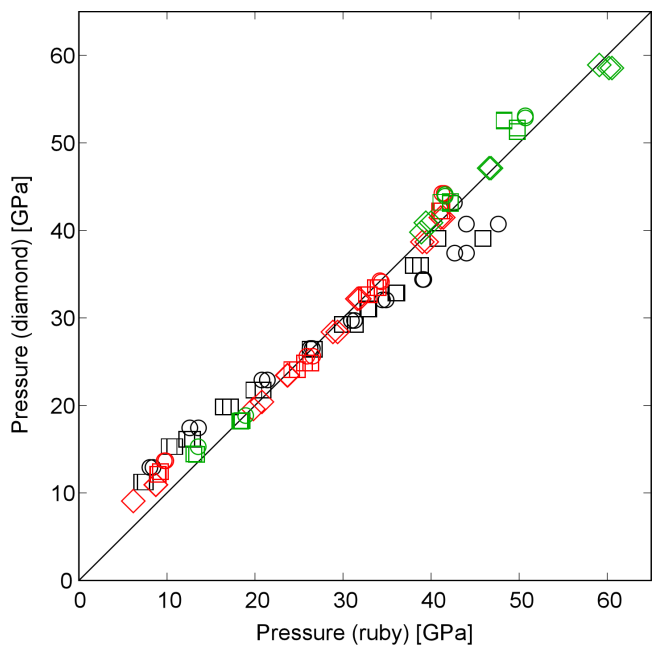

Supplementary Figure S4. Comparison of pressure determined by the diamond Raman gauge and the ruby florescence method. Values are given in Supplementary Tables S2 and S3. Black, red, and green symbols are for Runs 1, 2, and 3, respectively. Circles, squares, and diamonds are for 10, 120, and 300 K, respectively. The black line indicates $P$(ruby) = $P$(diamond).